# A nation's foreign and domestic professors: which have better research performance? (The Italian case)[1]

Giovanni Abramo

*Laboratory for Studies in Research Evaluation*
*at the Institute for System Analysis and Computer Science (IASI-CNR)*
*National Research Council of Italy*
ADDRESS: Istituto di Analisi dei Sistemi e Informatica, Consiglio Nazionale delle Ricerche, Via dei Taurini 19, 00185 Roma - ITALY
giovanni.abramo@uniroma2.it

Ciriaco Andrea D'Angelo

*University of Rome "Tor Vergata" - Italy and*
*Laboratory for Studies in Research Evaluation (IASI-CNR)*
ADDRESS: Dipartimento di Ingegneria dell'Impresa, Università degli Studi di Roma "Tor Vergata", Via del Politecnico 1, 00133 Roma - ITALY
dangelo@dii.uniroma2.it

Flavia Di Costa

*Research Value s.r.l.*
ADDRESS: Research Value, Via Michelangelo Tilli 39, 00156 Roma- ITALY
flavia.dicosta@gmail.com

## Abstract

This work investigates the research performance of foreign faculty in the Italian academic system. Incoming professors compose 1% of total faculty across the sciences, although with variations by discipline. Their scientific performance measured over 2010-2014 is on average better than that of their Italian colleagues: the greatest difference is for associate professors. Psychology is the discipline with the greatest concentration of top foreign scientists. However there are notable shares of unproductive foreign professors or of those with mediocre performance. The findings stimulate reflection on issues of national policy concerning attractiveness of the higher education system to skilled people from abroad, given the ongoing heavy Italian brain drain.

---

## 1. Introduction

Many scholars have illustrated that migration of skilled personnel is a critical aspect in knowledge-driven economies (She & Wotherspoon, 2013; Alfred, 2010; Smetherhama, Fentonb, & Modood, 2010; Kim, 2009; OECD, 2008; Universities UK, 2007; Sastry, 2005). There is increasing competitiveness for human capital, which in turn is a contributing factor in the cross-border movement of scholars, professors and experts (Knight, 2008).

Although the sciences are to some extent subject to national curricula, regulations, and administrative structures, they are globalised by definition, and therefore represent a specific case. In fact among developed countries, the development of research networks with inter-country and inter-sectoral mobility has become one of the top research policy objectives (OECD, 2008).

From the perspective of a specific country, mobility of skilled personnel takes place in two directions. The label "brain drain" suggests the outflow of a country's educated elite, to such extent as to menace the nation's long-term needs for development. The mirror-image term is "brain gain", indicating the results from programs, projects and other mechanisms for attraction of scientists to a given country (Jałowiecki & Gorzelak, 2004). In studying issues of intra-EU mobility in the highly qualified labor force, Giousmpasoglou and Koniordos (2017) specify that brain drain and gain refer to long-term or permanent emigration/immigration of highly qualified staff.

Gaillard & Gaillard (1998) instead conceived of "brain circulation" as a continuous flow in the skilled workforce, in a type of mobility between host and origin countries that stimulates the creation, diffusion and further adaptation of new knowledge. In effect, individual scientists can "design" different models of migration (permanent, temporary, circular) to individual or multiple countries over a period of time. This creates a complex model of spatial-temporal mobility (Ackers & Gill, 2008).

The 2010 OECD Innovation Strategy states that policies on mobility should aim to support knowledge flows and create enduring cross-country linkages and networks, enabling movement on a short-term or circular basis (OECD, 2010). However, as Van Noorden (2012) observes: "Science may increasingly be a globalized enterprise, but until would-be competitors boost their spending on science and facilities, it will simply give scientists even more opportunities to clump inside the countries that are already at the top of the pack". What occurs in reality is that the developed countries attract the highest proportions of foreign scientists, as shown the GlobSci survey conducted by Franzoni, Scellato and Stephan (2012). For almost all the 16 "core countries" surveyed, the United States is the dominant destination of emigration. While Swiss and Indian scientists are the most mobile, Americans are least likely to move. In countries such as India, Japan, Brazil and Spain, the number of foreign scientists and engineers is marginal, with percentages consistently below 10% of total. For Italy, these values drop to around 3%, with heavy weighting towards incoming scientists from France, Germany and Spain (respectively 13.0%, 11.1% and 11.1% of arrivals).

A recent study by Sugimoto et al. (2017), observing changes in the affiliations of authors for papers published over 2008-2015 (14 million publications in all), shows that Europe and Asia have in fact registered net losses of researchers (-22% and -20%), unlike the large gains recorded for North America (nearly +50%). It was also observed that most scientists continued to maintain links with their original countries, and that many had returned, confirming the appropriateness of the "brain circulation" concept in characterizing the current context, rather than brain drain or gain.

Concerning the USA, the 2016 National Science Foundation-Science and Engineering Indicators (National Science Board, 2016) resoundingly confirm the strength of incoming movement in the science and engineering fields: in 2013, foreign-born individuals[2] accounted for 27% of college-educated workers in these occupations, which is higher than their representation in both overall population (13%) and among college graduates in general (15%).

In Europe the tendency is towards growing intra-EU mobility of researchers (Lowell, 2007; Parey & Waldinger, 2011), thanks in part to EU policies which have engaged specific initiatives in favour of an integrated research area (e.g. Marie Curie fellowships). For Italy, a study by Todisco, Brandi, and Tattolo (2003), on foreign researchers working in public research institutions in 2001, confirmed the GlobSci survey, quantifying such scientists as a relatively low 4% of the active workforce. The objectives of the 2001 study were to analyze their socio-demographic characteristics, types of employment, duration of stay in Italy, and reasons for arriving and returning home. The literature on the incentives to mobility in science is in fact very rich, and identifies several strong drivers: wage premiums, career advancement and research opportunities, research facilities, the opportunity to work with significant peers and in prestigious institutions, and increased autonomy and freedom to debate and carry out research (Appelt, van Beuzekom, Galindo-Rueda, & de Pinho, 2015). Extra-professional motives, such as family and personal considerations, also come into play in scientists' migration and residency choices (OECD, 2010).

It the Italian case, it is precisely the structural lack of incentives, or "factors of appeal" that would explain the scarce number of incoming foreign researchers. As we will better explain in the next section of this paper, the overall legislative-administrative context has created a culture that is completely non-competitive, yet flourishing with favoritism and other opportunistic behaviors that are dysfunctional to the social and economic roles of the higher education system (Abramo, D'Angelo, & Rosati, 2016, 2015, 2014; Gerosa, 2001; Aiuti, Bruni, & Leopardi, 1994; Amadori, Bernasconi, Boccadoro, Glustolisi, & Gobbi, 1992). The overall result is a system of universities that are almost completely undifferentiated for quality and prestige (with a few exceptions), and unable to attract significant numbers of talented foreign faculty (Franzoni, Scellato, & Stephan, 2012; Todisco, Brandi, & Tattolo, 2003). With this study, the authors propose to examine further aspects of the issue, responding to the following research questions:

- What is the research performance of foreign professors, compared to that of their Italian colleagues in Italian universities?
- Are there differences across academic ranks and disciplines?

The aim is to verify whether Italy is able to attract foreign professors who can add higher value than the incumbents on the advancement of science. The analysis by discipline should reveal whether are more attractive than others. In the next section we will explore the background of the analysis and the main methodological issues. Section 3 presents the analytical results, while Section 4 offers the conclusions with authors' comments.

## 2. Background and method

[2] Meaning both long-term residents, well inserted in the context of the host country, and recent immigrants who continue to maintain strong social, educational, and economic links with their country of origin.

## 2.1 The Italian higher education system

The Italian Ministry of Education, Universities, and Research (MIUR) designates a total of 96 universities throughout Italy with the authority to issue legally recognized degrees. Of these 29 are small, private, special-focus universities (of which 13 offer only e-learning), and 67 are public and generally multi-disciplinary universities. Six of the whole are *Scuole Superiori* (Schools for Advanced Studies), specifically devoted to highly talented students, with very small faculties and tightly limited enrolment per degree program.

In the overall system, 94.9% of faculty are employed in public universities, and only 0.5% in *Scuole Superiori*. Public universities are largely financed by the government (56% of total income). Until 2009 the core government funding was input oriented (i.e. independent of merit, distributed to universities in a manner intended to equally satisfy the needs of each and all, varying only in respect to size and research disciplines). It was only following the first national research evaluation exercise, conducted in the period 2004-2006, that the MIUR began to assign a minimal share, equivalent to 4% of total income, on the basis of research and teaching assessment. The merit-based share has since increased to 20%.

Despite interventions intended to grant increased autonomy and responsibilities to the universities (Law 168 of 1989),[3] the Italian higher education system is a long-standing, classic example of a public and highly centralized governance structure, with low levels of autonomy at the university level and a very strong role played by the central state.

In keeping with the Humboldtian model, there are no "teaching-only" universities in Italy, as all professors are required to carry out both research and teaching. National legislation includes a provision that each faculty member must provide a minimum of 350 hours per year of instruction (including teaching, preparation to teaching, exams, thesis supervision, etc.). At the close of 2017, there were 56,000 faculty members in Italy (full, associate and assistant professors) and a roughly equal number of technical-administrative staff. Salaries are regulated at the central level and are calculated according to role (administrative, technical or professorial), rank within role (e.g. assistant, associate or full professor) and seniority. None of a professor's salary depends on merit. Moreover, as in all Italian public administration, dismissal of unproductive employees is unheard of. All new personnel enter the university system through public competitions, and career advancement depends on further public competitions.

New transparency provisions, the nomination of a national committee of experts in the field, and timely issue of regulations for the evaluation procedures, are all intended to ensure efficiency in the faculty selection process. In reality, the systemic characteristics – including a historically strong inclination to favoritism, structured absence of responsibility for poor performance by research units, and lack of merit incentive schemes – undermine the credibility of selection procedures for hiring and advancement of university personnel. This lack of credibility is accentuated in the public eye by the high and growing number of legal cases brought by unsuccessful candidates, by continual critical reports in the social and mass media, and by specific studies of systemic problems (Perotti, 2008; Zagaria, 2007).

[3] This law was intended to grant increased autonomy and responsibility to the universities to establish their own organizational frameworks, including charters and regulations. Subsequently, Law 537 (Article 5) of 1993 and Decree 168 of 1996 provided further changes intended to increase university involvement in overall decision-making on use of resources, and to encourage individual institutions to operate on the market and reach their own economic and financial equilibrium.

The overall result is that of a system of universities almost completely undifferentiated for quality and prestige, with the exception of the tiny *Scuole Superiori* and a very small number of private special-focus universities. Thus, also given the widespread knowledge of this context, the system is unable to attract significant numbers of talented foreign faculty, or even students. This is a system where: i) every university has some share of top scientists, flanked by another share of absolute non-producers (Abramo, Cicero, & D'Angelo, 2013a); ii) 23% of professors alone produce 77% of the overall Italian scientific advancement; iii) this 23% of 'top' faculty is not concentrated in a limited number of universities, but dispersed more or less uniformly among all Italian universities, along with the unproductive academics, so that no single institution reaches the critical mass of excellence necessary to develop as an elite university and to compete internationally (Abramo, Cicero, & D'Angelo, 2012a).

## 2.2 Measuring research performance

Research activity is a production process in which the inputs consist of human resources and other tangible (scientific instruments, materials etc.) and intangible (accumulated knowledge, social networks, reputation etc.) resources, and where outputs have a complex character of both tangible (publications, patents, conference presentations, databases, protocols etc.) and intangible natures (tacit knowledge, consulting activity etc.). Thus, the new-knowledge production function linking outputs to inputs, has a multi-input and multi-output character. The principal indicator of the efficiency of any production system is labor productivity.

The calculation of labor productivity requires a few simplifications and assumptions. It has been shown (Moed, 2005) that in the sciences, the prevalent form of codification of research output is publication in scientific journals. As a proxy of total output, in this work we consider only publications (articles, article reviews and proceeding papers) indexed in the Web of Sciences (WoS), leaving aside patents,[4] databases, and other forms of codification of new knowledge.

When measuring labor productivity, if there are differences in the production factors available to each scientist then one should normalize for them. Unfortunately, relevant data are not available at the individual level in Italy. The first assumption is that the same resources are available to all professors within the same field. The second assumption is that the hours devoted to research are more or less the same for all professors. Given the characteristics of the Italian academic system, as noted in Section 2.1, the above assumptions appear to be acceptable.

Most bibliometricians define productivity as the number of publications in the period of observation. Because publications have different values (impact), we prefer to adopt a more meaningful definition of productivity: the value of output per unit value of labor, all other production factors being equal. The latter recognizes that the publications embedding new knowledge have a different value (impact) on scientific advancement, which bibliometricians approximate with citations or journal impact factors. Provided that there is an adequate citation window (at least two years) the use of citations is always preferable (Abramo, Cicero, & D'Angelo, 2011). Because citation behavior varies by field, we standardize the citations for each publication with respect to the average of the distribution of citations for all the Italian publications indexed in the same year and the same WoS subject category.[5] Furthermore, research projects frequently involve a team of scientists, which is registered in the co-authorship of publications. In this case we account for the fractional contributions of scientists to outputs, which is sometimes further signaled by the position of the authors in the list of authors.

[4] Patents are often followed by publications that describe their content in the scientific arena, so the analysis of publications alone may actually avoid in many cases a potential double counting.

As performance indicator of professors we take the average yearly labour productivity, termed fractional scientific strength (*FSS*),[6] over the period 2010-2014.[7]

The indicator is defined as:

$$FSS = \frac{1}{t}\sum_{i=1}^{N}\frac{c_i}{\bar{c}}\mathrm{f_i} \qquad [1]$$

where:

$t$ = number of years of work by the researcher in period under observation
$N$ = number of publications by the researcher in period under observation
$c_i$ = citations received by publication *i* (counted on June 30, 2017)
$\bar{c}$ = average of distribution of citations received for all cited publications in same year and subject category of publication *i*
$\mathrm{f_i}$ = fractional contribution of researcher to publication *i*.

The fractional contribution equals the inverse of the number of authors in those fields where the practice is to place the authors in simple alphabetical order but assumes different weights in other cases. For Biology and Medicine, widespread practice in Italy is for the authors to indicate the various contributions to the published research by the order of the names in the listing of the authors. For Biology and Medicine, we thus give different weights to each co-author according to their position in the list of authors and the character of the co-authorship (intra-mural or extra-mural), as suggested in Abramo, D'Angelo, and Rosati (2013).[8]

Because of the differences in the intensity of publications across fields, a prerequisite of any distortion-free comparative performance assessment is to classify each researcher into a single field (Abramo, Cicero, & D'Angelo, 2013b).

---

[5] Abramo, Cicero, and D'Angelo (2012b) demonstrated that the average of the distribution of citations received for all cited publications of the same year and subject category is the best-performing scaling factor.

[6] An extensive theoretical dissertation on how to operationalize the measurement of productivity can be found in Abramo and D'Angelo (2014a).

[7] A five-year publication period is adequate to reduce the problem of paucity of publications at field level and year-dependent fluctuations with systematic effects on results (Abramo, D'Angelo, & Cicero, 2012).

[8] If first and last authors belong to the same university, 40% of citations are attributed to each of them; the remaining 20% are divided among all other authors. If the first two and last two authors belong to different universities, 30% of citations are attributed to first and last authors; 15% of citations are attributed to second and last but one author; the remaining 10% are divided among all others. The weighting values were assigned following advice from senior Italian professors in the life sciences. The values could be changed to suit different practices in other national contexts.

### 2.3 Data

Data on the faculty at each university were extracted from the database on Italian university personnel, maintained by the MIUR. For each professor this database provides information on their gender, affiliation, field classification and academic rank, at close of each year.[9]

In the Italian university system all academics are classified in one and only one field, named scientific disciplinary sector (SDS), 370 in all. SDSs are grouped into disciplines, named university disciplinary areas (UDAs), 14 in all. A key methodological issue concerns the limitation of the analysis to those fields where bibliometric analysis can be considered meaningful, i.e. publications indexed in bibliometric repositories (such as WoS or Scopus) provide a significant representation of the overall research output. We limit our field of analysis to the sciences, where the WoS coverage is acceptable for bibliometric assessment. Within the sciences, the analysis excludes those SDSs where less than 50% of professors achieve production of at least one WoS-indexed publication in the period under observation. The dataset thus formed consists of professors from 10 UDAs (mathematics and computer sciences, physics, chemistry, earth sciences, biology, medicine, agricultural and veterinary sciences, civil engineering, industrial and information engineering, psychology) and 203 SDSs. Again for reasons of significance, the analysis was limited to those professors (35,875 in all) who held formal faculty roles in these SDSs for at least three years over the 2010-2014 period.

Information on the citizenship of professors is not directly available. For identification of foreign professors we proceed by several steps and assumptions. First, the assignment of Italian tax codes permits identification of 757 foreign-born professors. Among these, we consider all those with both non-Italian name and family name as immigrants, and all those with both Italian name and family name as non-immigrants. Finally, for all those with Italian name/foreign family name or foreign name/Italian family name, we analyzed their curriculum vitae: among these we indicate those who had followed their educational path abroad as “incomers”. The final subset of foreign professors consists of 350 observations, equivalent to 1% of the 35,875 total professors in the dataset. The overall final dataset is presented in Table 1, by UDA (in brackets the share of foreign professors).

The bibliometric dataset used to measure *FSS* is extracted from the Italian Observatory of Public Research, a database developed and maintained by the present authors and derived under license from the Clarivate Analytics WoS Core Collection. Beginning from the raw data of the WoS, and applying a complex algorithm to reconcile the author’s affiliation and disambiguation of the true identity of the authors, each publication (article, review and conference proceeding) is attributed to the university scientist or scientists that produced it (D’Angelo, Giuffrida, & Abramo, 2011).[10] Thanks to this algorithm we can produce rankings of research productivity at the individual level, on a national scale. Based on the value of *FSS* we obtain a ranking list expressed on a percentile scale of 0-100 (worst to best) of all Italian academics of the same academic rank and SDS.

---

[9] http://cercauniversita.cineca.it/php5/docenti/cerca.php, last accessed on June 21, 2018.

[10] The harmonic average of precision and recall (F-measure) of authorships, as disambiguated by the algorithm, is around 97% (2% margin of error, 98% confidence interval).

***Table 1: Dataset of the analysis. Professors in the Italian academic system. In brackets the share of foreign professors***

| | | Professors | | | |
|---|---|---|---|---|---|
| UDA* | N. of SDSs | Assistant | Associate | Full | Total |
| 1 | 10 | 926 (1.8%) | 1,247 (1.9%) | 1,041 (3.5%) | 3,214 (2.4%) |
| 2 | 8 | 573 (0.7%) | 1,055 (1.6%) | 652 (0.9%) | 2,280 (1.2%) |
| 3 | 12 | 920 (0.5%) | 1,263 (0.9%) | 750 (0.7%) | 2,933 (0.7%) |
| 4 | 12 | 353 (0.6%) | 454 (1.3%) | 278 (0.7%) | 1,085 (0.9%) |
| 5 | 19 | 1,880 (0.9%) | 1,723 (1.4%) | 1,280 (0.5%) | 4,883 (1.0%) |
| 6 | 50 | 4,253 (0.8%) | 3,401 (0.6%) | 2,442 (0.2%) | 10,096 (0.6%) |
| 7 | 30 | 1,034 (1.0%) | 1,135 (1.2%) | 868 (0.5%) | 3,037 (0.9%) |
| 8 | 10 | 471 (1.3%) | 621 (0.5%) | 512 (0.0%) | 1,604 (0.6%) |
| 9 | 42 | 1,377 (1.2%) | 2,110 (0.7%) | 1,777 (0.6%) | 5,264 (0.8%) |
| 10 | 10 | 487 (2.3%) | 514 (2.5%) | 388 (1.5%) | 1,389 (2.2%) |
| Total | 203 | 12,274 (1.0%) | 13,523 (1.1%) | 9,988 (0.8%) | 35,785 (1.0%) |

** 1, Mathematics and computer science; 2, Physics; 3, Chemistry; 4, Earth sciences; 5, Biology; 6, Medicine; 7, Agricultural and veterinary sciences; 8, Civil engineering; 9, Industrial and information engineering; 10, Psychology*

## 3. Analysis

Given the above assumptions concerning identification of foreign professors, their presence in the Italian academic-scientific disciplines results as truly minimal, being limited to 1% of the total. Their distribution is quite uniform among academic ranks, with a very slight prevalence for associates (1.1%) and a lesser concentration among full professors (0.8%). The distribution among UDAs is more differentiated: in mathematics and computer science, foreign professors are 2.4% of total, with a significant concentration in the top role of full professor (36 out of 1041, i.e. 3.5%). In psychology and physics the share of foreign professors is again greater than average, at 2.2% and 1.2% respectively. In all the other UDAs foreign professors represent less than 1% of total, with the minimum (0.6%) seen in medicine and in civil engineering; in this latter UDA there are also no observations of foreign full professors.

The foreign professors come from 59 countries (Figure 1 and Figure 2): the greatest frequency is observed for Germany (35), followed by the United States (26) and Greece (18). Of the 350 foreign professors, 190 are from EU-28 countries. There are 32 professors from Asian countries (eight from Iran and six from China), 17 from Africa (nine from the so-called Maghreb countries).[11] Thirty-five professors are from Latin America, with Argentina most represented (13). From Oceania there are six professors: five Australian and one New Zealander.

Given the dataset, we can now respond to the research questions, analyzing the research performance of the 350 foreign professors. Table 2 shows the average percentile value of FSS registered for those falling in each UDA. The last row reports an overall value of 56.5, slightly higher than the median (50). The maximum (63.7) is seen for the 48 professors of biology (UDA 5), followed by those of medicine (59.2) and psychology (58.8). Below the “expected” value of 50 we have physics (46.9) and chemistry (48.4). In physics in particular Italy has a long and important tradition of international standing, as shown by the Italian Nobel prize laureates (Guglielmo Marconi, Enrico Fermi, Emilo Segrè, just to cite a few). Furthermore, physics outstands all other disciplines by incidence of professors authoring highly cited articles (Abrano and D’Angelo, 2014c; ANVUR, 2016). The resulting attractiveness of physics is shown by the average performance of young talented foreign assistant professors hired in Italian universities, ten points above that of their Italian peers. On the contrary, foreign full professors find it hard to compete (45.7%) against  long-experienced Italian peers.

[11] Tunisia, Algeria and Morocco.

Analyzing the populations belonging to each academic rank, again for average percentile of FSS, we see an absolute maximum of 79.8 for the four full professors in agricultural and veterinary sciences and a minimum of 31.5 for the two assistant professors of earth sciences. In general, it is the assistant professors with the lowest overall average percentile of FSS, at 54.7, compared to 58.1 for the associates and 56.5 for full professors. In physics instead the average performance of foreign assistant professors is better than that of associate and full professors; the same holds true also in psychology.

***Figure 1: Provenience of foreign professors, global data***

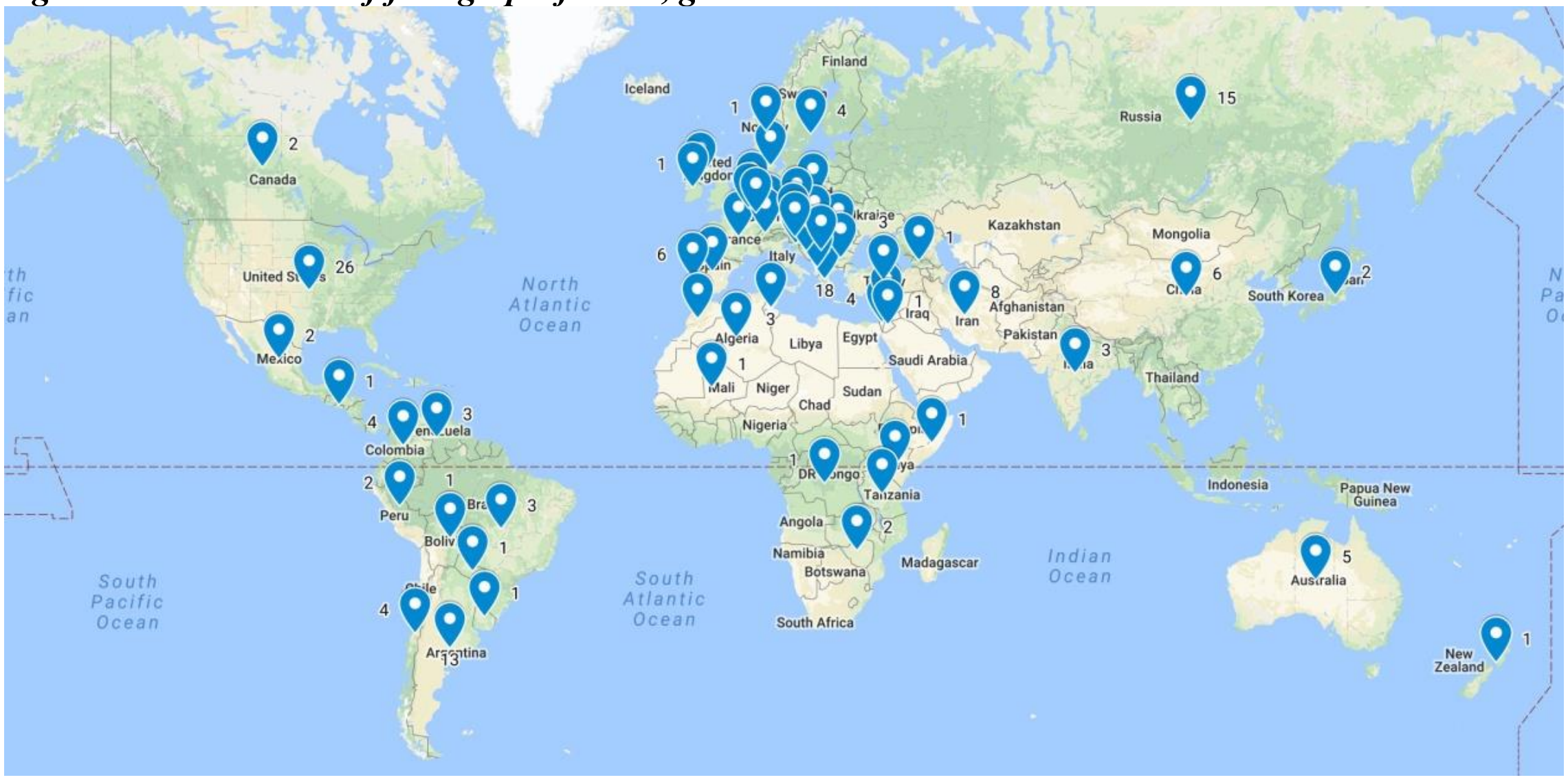

*Figure 2: Provenience of foreign professors, from European country*

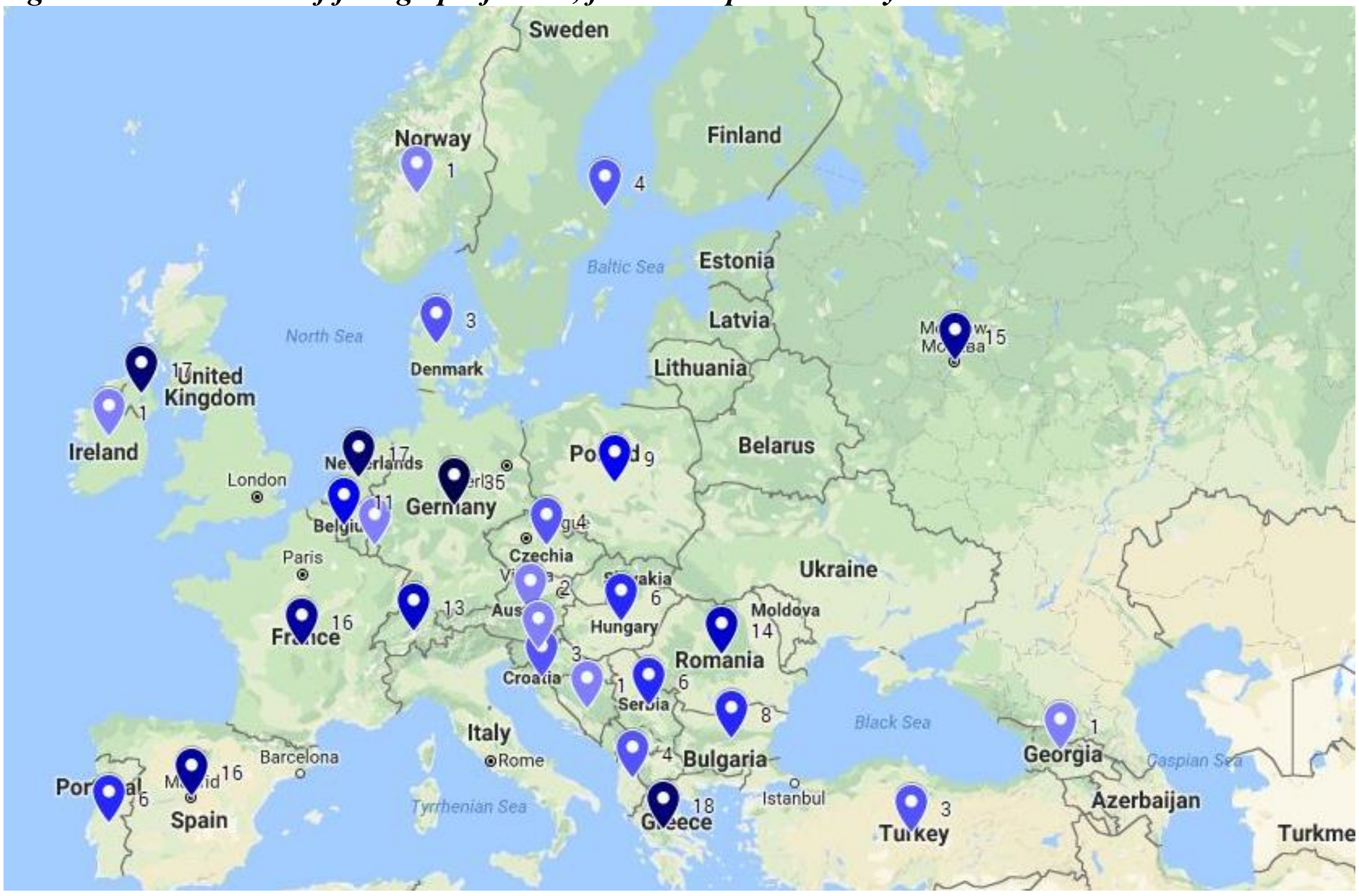

*Table 2: Average FSS percentile for foreign professors, by UDA and academic rank*

| | Assistant professors | | Associate professors | | Full professors | | Total | |
|---|---|---|---|---|---|---|---|---|
| UDA* | Obs | Average FSS | Obs | Average FSS | Obs | Average FSS | Obs | Average FSS |
| 1 | 17 | 49.7 | 24 | 57.4 | 36 | 53.4 | 77 | 53.8 |
| 2 | 4 | 60.1 | 17 | 44.2 | 6 | 45.7 | 27 | 46.9 |
| 3 | 5 | 34.7 | 11 | 58.4 | 5 | 40.0 | 21 | 48.4 |
| 4 | 2 | 31.5 | 6 | 60.4 | 2 | 49.5 | 10 | 52.4 |
| 5 | 17 | 57.9 | 24 | 65.5 | 7 | 71.9 | 48 | 63.7 |
| 6 | 35 | 57.7 | 20 | 61.8 | 5 | 60.0 | 60 | 59.2 |
| 7 | 10 | 54.8 | 14 | 54.4 | 4 | 79.8 | 28 | 58.1 |
| 8 | 6 | 52.4 | 3 | 69.2 | 0 | n.a | 9 | 58.0 |
| 9 | 16 | 52.1 | 14 | 55.8 | 10 | 69.4 | 40 | 57.7 |
| 10 | 11 | 64.2 | 13 | 61.2 | 6 | 43.6 | 30 | 58.8 |
| Total | 123 | 54.7 | 146 | 58.1 | 81 | 56.5 | 350 | 56.5 |

** 1, Mathematics and computer science; 2, Physics; 3, Chemistry; 4, Earth sciences; 5, Biology; 6, Medicine; 7, Agricultural and veterinary sciences; 8, Civil engineering; 9, Industrial and information engineering; 10, Psychology*

Use of the average as indicator of central tendency is open to criticism, considering the low number of observations. To respond, in Table 3, we report the percentage of foreign professors in selected FSS distribution tiles. Examining the right tail of the performance distribution, meaning top scientists, the values are clearly higher than expected. The top 1% (foreign scientists with FSS percentile not less than 99) represent 4% of total; lowering the threshold to the 95th percentile, we arrive at 8% of total; finally, examining the first decile by performance, foreign professors represent a near 15% share. At the UDA level, the data for psychology are striking: three of 30 foreign professors (meaning 10% of total) are absolute top. In industrial and information engineering, and in mathematics and computer science, the incidence of top 1% scientists is again high, at 7.5% and 6.5% respectively, followed by chemistry (4.8%) and biology (4.2%). In the other UDAs we do not observe any top 1% professors, however among the top 5% we still note shares greater than expected: in physics (7.4%), in earth sciences (10.0%) and in civil engineering (11.1%). Finally, among the top 10%, only physics and chemistry register a share of foreign professors less than the expected values.

***Table 3: Share of foreign professors in each FSS distribution tile, by UDA***

| UDA* | Obs | Top 99% | Top 95% | Top 90% | Top 50% | Bottom 20% | Unproductive |
|---|---|---|---|---|---|---|---|
| 1 | 77 | 6.5% | 10.4% | 20.8% | 57.1% | 16.9% | 11.7% |
| 2 | 27 | 0.0% | 7.4% | 7.4% | 55.6% | 29.6% | 3.7% |
| 3 | 21 | 4.8% | 4.8% | 9.5% | 47.6% | 23.8% | 0.0% |
| 4 | 10 | 0.0% | 10.0% | 10.0% | 50.0% | 30.0% | 0.0% |
| 5 | 48 | 4.2% | 8.3% | 16.7% | 68.8% | 8.3% | 0.0% |
| 6 | 60 | 0.0% | 1.7% | 10.0% | 63.3% | 6.7% | 3.3% |
| 7 | 28 | 0.0% | 3.6% | 14.3% | 64.3% | 14.3% | 3.6% |
| 8 | 9 | 0.0% | 11.1% | 11.1% | 55.6% | 11.1% | 11.1% |
| 9 | 40 | 7.5% | 10.0% | 12.5% | 65.0% | 12.5% | 5.0% |
| 10 | 30 | 10.0% | 16.7% | 23.3% | 63.3% | 16.7% | 16.7% |
| Total | 350 | 4.0% | 8.0% | 14.9% | 60.9% | 14.9% | 6.0% |

** 1, Mathematics and computer science; 2, Physics; 3, Chemistry; 4, Earth sciences; 5, Biology; 6, Medicine; 7, Agricultural and veterinary sciences; 8, Civil engineering; 9, Industrial and information engineering; 10, Psychology.*

The percentage of foreign professors with performance greater than the median value is 60.9%, more than 10 percentage points greater than the expected value. At the UDA level the data vary from a minimum of 47.6% in chemistry to a maximum of 68.8% in biology. However there is also a notable presence of foreign professors in the left tail of the performance distribution: almost 15% place in the last quintile for performance, with a peak of a full 30% in earth sciences, followed by substantial shares in physics (29.6%) and chemistry (23.8%). Among the unproductive professors (FSS = 0), the highest share of foreigners appears in psychology: three of five unproductive foreign professors did not publish any works indexed in the WoS, while two achieved publications which were then never cited. In chemistry, earth sciences and biology there were no unproductive foreign professors.

We remind that the above results and analysis may suffer from all the usual limits, caveats, and assumptions of evaluative scientometrics, in particular: i) publications are not representative of all knowledge produced; ii) bibliometric repertories do not cover all publications; iii) citations are not always certification of real use and representative of all use; and iv) capital is assumed to be equal. A thorough discussion of the basic concepts and limits of evaluative scientometrics can be found in Abramo (2018). Furthermore, the identification of foreign professors is not certain, as the relevant data are not directly available.

## 4. Conclusions

The phenomenon of the Italian brain drain receives intense media attention in the home country. Not a day goes by without a phalanx of media and social outlets offering pessimistic stories of the latest brilliant young researchers, deluded by their home country, gone to find fortune elsewhere. “Brain drain” (*fuga dei cervelli*) has become an integral part of the popular lexicon, one of the most repeated expressions in the political discourse, heard from all parts of the political spectrum. Some critics go so far as to speak of an "exodus" comparable to Italian mass migration of the 19th century, with the difference that today's leavers are above all the "skilled people" in whom the country has invested heavily, from primary to higher education.

The *Virtual Italian Academy*, a network association of Italian academics and professionals throughout Europe, in its census of Italian top scientists,[12] traced 780 of these operating abroad in the science fields, of which 682 were completely without ties to any Italian institution, and 325 were based in the USA.[13] This particular census dealt only with "top scientists", meaning scarcely touching on the whole migration phenomenon of scholars.

The supposed brain drain would raise less outrage if it were in some way accompanied by demonstration of a compensating phenomenon of similar scope: of flows in the opposite incoming direction, meaning "brain gain". With this in mind, the intention of the current study was to investigate the dimensions of the foreign faculty presence in the Italian academic system, and then the research performance of the foreign tenure-track professors operating in the nation's universities, as compared to Italian professors. The study confirms the data of preceding analyses, which indicated a truly minimal representation of foreign researchers working in Italy (Todisco, Brandi, & Tattolo, 2003). In the sciences, foreign tenure-track professors represent only 350 of 35,800 total professors operating in Italian universities, less than 1% of the total faculty. There are disciplinary variations: in mathematics and computer science, foreigners were 2.4% of total; in medicine and in civil engineering the "foreign share" dropped to 0.6%. Half of the foreign professors are from other EU-28 countries, with Germany in the lead. Twenty-six professors are from the USA.

Meanwhile, viewing from the other side, in 2011 there were 110,000 incoming scholars working at colleges and universities in the United States (Foderaro, 2011) - as researchers, instructors and professors – and at the same time, not less than 15% of American universities have foreign-born deans.

Regarding the ability of the Italian academic system to attract numbers of foreign professors, the analytical data paint a glaringly obvious picture. However the judgment on the scientific performance of incoming professors remains suspended or mixed: the productivity indicator calculated on the 2010-2014 scientific production indexed in WoS reveals that foreign faculty are on average better than their Italian colleagues, in particular the associate professors. For the assistants the performance advantage is less. Foreign professors in the bio-medical area are particularly capable. In physics and in chemistry the average performance of incomers is instead less than that of their Italian colleagues, perhaps because Italian has a long-standing tradition of excellence in these two disciplines. Psychology is the discipline with the greatest concentration of foreign top scientists: the analysis shows that in this discipline, the top 1%, 5% and 10% tiles contain more than double the expected share. Mathematics shows a similar situation. Still, there are also notable shares of foreign professors with poor to mediocre performance: 15% of foreign professors fall in the bottom quintile of performance, with a peak of 30% in earth sciences.

Science is by definition an open, cosmopolitan, globalised system. Although each country will always have characteristic features, for the developed countries in particular, a "healthy" research sphere would demonstrate a condition of "brain circulation", in which the flows entering and exiting are at least partially self-compensating. In reality, as for all migration, the relative levels of flows are a function of differences in potential, in attractivity, leading to more or less imbalanced movements from or towards certain countries. It is on the fact of this “differential” that the policy maker is called to act, to prevent that the outflows of the country’s educated elite, unaccompanied by compensating arrivals, become so great as to menace long-term national development.

[12] Defined as such on the basis of an h index not less than 30 calculated on their overall scientific production indexed in Google Scholar.
[13] http://www.topitalianscientists.org, last accessed on June 21, 2018.

The Italian situation is clearly one of ongoing damage, at the very least of high risk, calling for responsible action from policy makers. This is not the first nor likely the last of the appeals launched by the authors in this regard. There is a need for assertive, even disruptive action in the governance of the overall system, providing greater autonomy to the universities, but at the same time demanding greater responsibility for the results achieved through the exercise of autonomy. There must be a conversion to truly competitive mechanisms, robust evaluation processes and adequate rewarding of merit, clarity and certainty in paths of career entry and progress, but also effective and certain sanctioning systems for opportunistic and patronage practices.

A further and still more courageous step for Italy would be to link the professors’ actual salaries to the quality of their teaching and research. Similar incentives have recently been introduced in the public administration sector, but not yet in higher education, which is itself an indication of the reigning conservative forces. There is room for further and much needed initiatives, which would be greatly effective in stimulating greater competition and continuous improvement: liberation of resources cutting unproductive faculty, and abolition of the legal value of undergraduate degrees.

Finally, no international university ranking lists an Italian university above 150$^{th}$ place. The absence of elite universities in Italy certainly affects the ability of the country to attract foreign talents and retain domestic ones. The most productive faculty are not concentrated in a limited number of universities, but is instead dispersed more or less uniformly among all Italian universities, along with the unproductive individuals, so that no single institution reaches the critical mass of excellence necessary to develop as an elite university and compete at the international level. The question is how rapidly achieving in Italy what competitive systems (see USA and UK) have produced over the span of decades. Abramo and D’Angelo (2014b) suggest an intervention that could proceed independently, and would still be complementary to the above said measures: an intervention fostering the birth of spin-off universities, newly staffed by migration of only the top scientists from the existing universities and research institutions. This intervention would be much less unpopular because it would not touch vested interests, and so would encounter less resistance to implementation.

All of these are directions that would return the Italian higher education system to its just social and economic role, boosting its attractiveness in the global context, thereby reducing the brain drain and increasing brain gain.

## References

Abramo, G. (2018). Revisiting the scientometric conceptualization of impact and its measurement. *Journal of Informetrics,* 12(3), 590-597.

Abramo, G., & D’Angelo, C.A. (2014a). How do you define and measure research productivity? *Scientometrics,* 101(2), 1129-1144.

Abramo, G., & D'Angelo, C.A. (2014b). The spin-off of elite universities in non-competitive, undifferentiated higher education systems: an empirical simulation in Italy. *Studies in Higher Education,* 39(7), 1270-1289.
Abramo, G., D'Angelo, C.A. (2014c). Assessing national strengths and weaknesses in research fields. *Journal of Informetrics,* 8(3), 766-775.
Abramo, G., Cicero, T., & D'Angelo, C.A. (2011). Assessing the varying level of impact measurement accuracy as a function of the citation window length. *Journal of Informetrics,* 5(4), 659-667.
Abramo, G., Cicero, T., & D'Angelo, C.A. (2012a). The dispersion of research performance within and between universities as a potential indicator of the competitive intensity in higher education systems. *Journal of Informetrics,* 6(2), 155-168.
Abramo, G., Cicero, T., & D'Angelo, C.A. (2012b). Revisiting the scaling of citations for research assessment. *Journal of Informetrics*, 6(4), 470-479.
Abramo, G., Cicero, T., & D'Angelo, C.A. (2013a). The impact of unproductive and top researchers on overall university research performance. *Journal of Informetrics,* 7(1), 166-175.
Abramo, G., Cicero, T., & D'Angelo, C.A. (2013b). Individual research performance: a proposal for comparing apples to oranges. *Journal of Informetrics*, 7(2), 528-539.
Abramo, G., D'Angelo, C.A., & Cicero, T. (2012). What is the appropriate length of the publication period over which to assess research performance? *Scientometrics,* 93(3), 1005-1017.
Abramo, G., D'Angelo, C.A., & Rosati, F. (2013). The importance of accounting for the number of co-authors and their order when assessing research performance at the individual level in the life sciences. *Journal of Informetrics,* 7(1), 198-208.
Abramo, G., D'Angelo, C.A., & Rosati, F. (2014). Career advancement and scientific performance in universities. *Scientometrics*, 98(2), 891-907.
Abramo, G., D'Angelo, C.A., & Rosati, F. (2015). The determinants of academic career advancement: evidence from Italy. *Science and Public Policy*, 42(6), 761-774.
Abramo, G., D'Angelo, C.A., & Rosati, F. (2016). Gender bias in academic recruitment. *Scientometrics*, 106(1), 119-141.
Ackers, H. L., & B. Gill. (2008). *Moving people and knowledge: Understanding the processes of scientific mobility within an enlarging Europe*. Cheltenham: Edward Elgar.
Aiuti, F, Bruni, R., & Leopardi, R. (1994). Impediments of Italian science. *Nature*, 367(6464), 590-590.
Alfred, M. V. (2010). Transnational migration, social capital and lifelong learning in the USA. *International Journal of Lifelong Education*, 29(2), 219-235.
Amadori, S., Bernasconi, C., Boccadoro, M., Glustolisi, R., & Gobbi, M. (1992). Academic promotion in Italy. *Nature*, 355(6361), 581-581.
ANVUR (2017). *Valutazione della qualità della ricerca 2010-2011. Rapporto finale*. Retrieved from http://www.anvur.it/rapporto-2016/, last accessed on June 21, 2018
Appelt, S., van Beuzekom, B., Galindo-Rueda, F., & de Pinho, R. (2015). Which factors influence the international mobility of research scientists? In Geuna, A. (Ed.). *Global mobility of research scientists: the economics of who goes where and why. New York*, NY: Academic Press, 177-214.
D'Angelo, C.A., Giuffrida C., & Abramo, G. (2011). A heuristic approach to author name disambiguation in bibliometrics databases for large-scale research assessments.

*Journal of the American Society for Information Science and Technology*, 62(2), 257-269.
Foderaro, L.W., (2011), More Foreign-Born Scholars Lead U.S. Universities, *New York Times*, March 9, www.nytimes.com/2011/03/10/education/10presidents.html, last accessed on June 21, 2018.
Franzoni, C., Scellato, G., & Stephan, P. (2012). Foreign-born scientists: mobility patterns for 16 countries. *Nature Biotechnology*, 30(12), 1250-1253.
Gaillard, A.M., & Gaillard, J. (1998). The international circulation of scientists and technologists. *Science Communication*, 20(1), 106-115.
Gerosa, M. (2001). Competition for academic promotion in Italy. *Lancet*, 357(9263), 1208-1208.
Geuna, A. (Ed.). (2015). *Global mobility of research scientists: the economics of who goes where and why*. New York, NY: Academic Press.
Giousmpasoglou, C., & Koniordos, S. K. (2017). Brain drain in higher education in Europe: current trends and future perspectives. http://eprints.bournemouth.ac.uk/28702/1/Chapter.ID_47546_6x9-1-edited.pdf, last accessed on June 21, 2018.
Jałowiecki, B., & Gorzelak, G. J. (2004). Brain drain, brain gain, and mobility: Theories and prospective methods. *Higher Education in Europe*, 29(3), 299-308.
Kim, T. (2009). Transnational academic mobility, internationalization and interculturality in higher education. *Intercultural education*, 20(5), 395-405.
Knight, J. (2008). *Higher education in turmoil. The Changing World of Internationalisation*. Rotterdam, The Netherlands: Sense Publishers.
Lowell, L. B. (2007). Trends in international migration flows and stocks, 1975-2007, OECD Social, Employment and Migration Working Papers (58).
Moed, H. F. (2005). *Citation Analysis in Research Evaluation*. Springer, The Netherlands.
National Science Board. 2016. Arlington, VA: National Science Foundation Science and Engineering Indicators 2016 (NSB-2016-1), Immigration and the S&E Workforce Immigration and the S&E Workforce (3-101).
OECD (2008). Global Competition for Talent, Mobility of the Highly Skilled. Paris: OECD.
OECD (2010). The OECD innovation strategy: Getting a head start on tomorrow. Paris: OECD.
Parey, M., & Waldinger, F. (2011). Studying abroad and the effect on international labor market mobility: evidence from the introduction of ERASMUS. *The Economic Journal*, 121(551), 194-222.
Perotti, R. (2008). *L'università truccata*. Einaudi, Torino, Italy. ISBN: 978-8-8061-9360-7.
Sastry, T. (2005). Migration of Academic Staff to and from the UK: An Analysis of the HESA Data. Oxford: Higher Education Policy Institute.
She, Q., & Wotherspoon, T. (2013). International student mobility and highly skilled migration: A comparative study of Canada, the United States, and the United Kingdom. *SpringerPlus*, *2*(1), 132.
Smetherham, C., Fenton, S., & Modood, T. (2010). How global is the UK academic labour market? *Globalisation, Societies and Education, 8*(3), 411-428.

Sugimoto, C. R., Robinson-Garcia, N., Murray, D. S., Yegros-Yegros, A., Costas, R., & Larivière, V. (2017). Scientists have most impact when they're free to move. *Nature News*, 550(7674), 29.

Todisco, E., Brandi, M. C., & Tattolo, G. (2003). Skilled migration: a theoretical framework and the case of foreign researchers in Italy. *Fulgor*, 1(3), 115-130.

Universities UK (2007). Talent Wars: The International Market for Academic Staff. https://web.archive.org/web/20120112220251/http://www.universitiesuk.ac.uk/Publications/Documents/Policy%20Brief%20Talent%20Wars.pdf, last accessed on June 21, 2018.

Van Noorden, R. (2012). Science on the move. *Nature*, 490(7420), 326.

Weinberg, B. A. (2011). Developing science: scientific performance and brain drains in the developing world. *Journal of Development Economics*, 95(1), 95-104.

Zagaria, C. (2007). *Processo all'università. Cronache dagli atenei italiani tra inefficienze e malcostume*. Dedalo, Bari, Italy. ISBN: 978-8-8220-5365-7.